\documentclass[12pt]{article}
\usepackage{amsmath}

\newcommand{\bm}{\begin{multiline}}
\newcommand{\beq}{\begin{equation}}
\newcommand{\eeq}{\end{equation}}
\newcommand{\beqs}{\begin{eqnarray}}
\newcommand{\eeqs}{\end{eqnarray}}

\newcommand{\ra}{\rightarrow}

\begin{document}

\thispagestyle{empty}

\begin{flushright}
hep-th/yymmxxx
\end{flushright}

\hfill{}

\hfill{}

\hfill{}

\vspace{32pt}

\begin{center}
\textbf{{\Large New Taub-NUT-Reissner-Nordstr\"{o}m spaces in higher
dimensions }} \\[0pt]

\vspace{48pt}
{\bf Robert Mann}\footnote{E-mail: 
{\tt mann@avatar.uwaterloo.ca}}
{\bf and Cristian Stelea}\footnote{E-mail: 
{\tt cistelea@uwaterloo.ca}}

\vspace*{0.2cm}

{\it $^{1}$Perimeter Institute for Theoretical Physics}\\
{\it 31 Caroline St. N. Waterloo, Ontario N2L 2Y5 , Canada}\\[.5em]

{\it $^{1,2}$Department of Physics, University of Waterloo}\\
{\it 200 University Avenue West, Waterloo, Ontario N2L 3G1, Canada}\\[.5em]
\end{center}

\vspace{48pt}

\begin{abstract}
We construct new charged solutions of the Einstein-Maxwell field equations
with cosmological constant. These solutions describe the nut-charged
generalisation of the higher dimensional Reissner-Nordstr\"{o}m spacetimes. For
a negative cosmological constant these solutions are the charged
generalizations of the topological nut-charged black hole solutions in
higher dimensions. Finally, we discuss the global structure of such
solutions and possible applications.
\end{abstract}

\setcounter{footnote}{0}

\newpage

\section{Introduction}

Despite their unphysical and bizarre properties, nut-charged spacetimes are
receiving increased attention in recent years. Intuitively a nut charge
corresponds to a magnetic type of mass. Its presence induces a so-called
Misner singularity in the metric, analogous to a `Dirac string' in
electromagnetism \cite{Misner}. This singularity is only a coordinate
singularity and can be removed by choosing appropriate coordinate patches.
However, expunging this singularity comes at a price: in general we must
make coordinate identifications in the spacetime that yield CTCs in certain
regions.

The presence of closed time-like curves (CTCs) in their Lorentzian section
is a less than desirable feature. However, one could argue that it is
precisely this feature that makes them more interesting. Recently the
nut-charged spacetimes have been used in string theory as testbeds for the
AdS/CFT conjecture \cite{Hawking, Chamblin1}. Another very interesting
result was obtained via a non-trivial embedding of the Taub-NUT geometry in
heterotic string theory, with a full conformal field theory definition (CFT) %
\cite{Johnson1}. It was found that nutty effects were still present even in
the exact geometry, computed by including all the effects of the infinite
tower of massive string states that propagate in it. This might be a sign
that string theory can very well live even in the presence of nonzero NUT
charge, and that the possibility of having CTCs in the background can still
be an acceptable physical situation. Furthermore, it has been recently noted
that the boundary metric Lorentzian sector of these spaces in
AdS-backgrounds is in fact similar with the G\"{o}del metric \cite%
{micky2,micky3}. In addition, in asymptotically dS settings, regions near
past/future infinity do not have CTCs, and NUT-charged asymptotically dS
spacetimes have been shown to yield counter-examples to some of the
conjectures advanced in the still elusive dS/CFT paradigm \cite{strominger}-
such as the maximal mass conjecture and Bousso's entropic N-bound conjecture %
\cite{rick1,rick2,rickreview}. Moreover, they have been used to uncover deep
results regarding the gravitational entropy \cite%
{Hawking,Hunter,Robinson,Akbar,Mann1,Lorenzo} and, in particular, they
exhibit breakdowns of the usual relation between entropy and area \cite%
{micky1} (even in the absence of Misner string singularities).

The first solution in four dimensions describing such an object was
presented in \cite{Taub,NUT}. There are known extensions of the Taub-NUT
solutions to the case when a cosmological constant is present and also in
the presence of rotation \cite{Demiansky,Gibbons:nf,Ortin}. In these
cosmological settings, the asymptotic structure is only locally de Sitter
(for a positive cosmological constant) or anti-de Sitter (for a negative
cosmological constant) and we speak about Taub-NUT-(a)dS solutions.

Generalisations to higher dimensions follow closely the four-dimensional
case \cite{Bais,Page,Awad,Klemm,csrm,Page1,Chen,Gauntlett,TNletter}. In
constructing these metrics the idea is to regard Taub-NUT spacetimes as
radial extensions of $U(1)$ fibrations over a $2k$-dimensional base space
endowed with an Einstein-K\"{a}hler metric $g_{B}$. Then the $(2k+2)$
-dimensional Taub-NUT spacetime has the metric: 
\begin{equation}
F^{-1}(r)dr^{2}+(r^{2}+N^{2})g_{B}-F(r)(dt+\mathcal{A})^{2}  \label{I1}
\end{equation}
where $t$ is the coordinate on the fibre $S^{1}$ and the one-form $\mathcal{A%
}$ has curvature $\mathcal{J}=d\mathcal{A}$, which is proportional to some
covariantly constant $2$-form. Here $N$ is the NUT charge and $F(r)$ is a
function of $r$. Recently NUT-charged spacetimes with more than one nut
parameter have been obtained as exact solutions to higher-dimensional
Einstein field equations \cite{csrm}. These solutions have been later
generalised to arbitrary dimensions \cite{TNletter}.

There also exists a nut-charged generalisation of the Reissner-Nordstr\"{o}m solution in four dimensions \cite{Ortin,Brill}. However, there are no known higher dimensional generalisations of the nut-charged solutions in Einstein-Maxwell theory. The main purpose of this paper is to provide the
generalisation of these spaces to higher dimensions. These solutions will represent the electromagnetic generalisation of the nut-charged spacetimes studied in refs. \cite{Page1,Bais,Robinson,Awad,Lorenzo} as well as the nut-charged generalisation of the higher dimensional Reissner-Nordstr\"{o}m solutions \cite{Chamblin}.

Einstein-Maxwell theory in $D$-dimensions is described by the following action: 
\begin{eqnarray}
I&=&-\frac{1}{16\pi G}\int d^{D}x\sqrt{-g}\big[R-2\Lambda-F^2\big]
\end{eqnarray}
The equations of motion derived from this action can be written in the
following form\footnote{We use here $\lambda=\pm\frac{D-1}{l^2}$.}: 
\begin{eqnarray}
R_{\mu\nu}&=&2\left(F_{\mu\rho}F_{\nu}^{~\rho}-\frac{1}{2(D-2)} F^2g_{\mu\nu}\right)+\lambda g_{\mu\nu}  \notag \\
\nabla_{\mu}F^{\mu\nu}&=&0
\end{eqnarray}
where $F=dA$ is the electromagnetic $2$-form field strength corresponding to
the gauge potential $A$. Our conventions are as follows: we take $(-,+,...,+)$ for the (Lorentzian)
signature of the metric; in even $D$ dimensions our metrics will be solutions of the sourceless Einstein-Maxwell field equations with cosmological constant $\Lambda =\pm \frac{(D-1)(D-2)}{2l^{2}}$.

\section{The general solution}

Let us recall first the form of the $4$-dimensional Taub-NUT-Reissner-Nordstr\"{o}m metric. The metric is given by: 
\begin{eqnarray}
ds^2&=&-F(r)(dt-2N\mathcal{A})^{2}+F^{-1}(r)dr^{2}+(r^{2}+N^{2})g_{M}  \label{I2}
\end{eqnarray}
where $M$ is a $2$-dimensional Einstein-K\"{a}hler manifold, which can be taken to be the unit sphere $S^2$, torus $T^2$ or the hyperboloid $H^2$. In each case we have:
\begin{eqnarray}
\mathcal{A}&=& \Biggl\{ 
\begin{array}{cl}
\cos\theta d\phi, & ~~~~\mathrm{{for~~ \delta=1~(sphere)}\nonumber} \\ 
\theta d\phi, & ~~~~\mathrm{{for ~~\delta=0~ (torus)}\nonumber} \\ 
\cosh\theta d\phi, & ~~~~\mathrm{for~~ \delta=-1~(hyperboloid)}%
\end{array}%
\end{eqnarray}
while the function $F(r)$ and the gauge field potential $A$ have the following expressions:
\beqs
F(r)&=&\frac{r^4+(l^2+6N^2)r^2-2ml^2r-3N^4+l^2(q^2-N^2)}{l^2(r^2+N^2)}\nonumber\\
A&=&-\frac{qr}{r^2+N^2}\left(dt-2N\mathcal{A}\right)
\eeqs
Here $m$, $q$ and $N$ are respectively the mass, charge and the NUT parameter. As one can see directly from the expression of the $1$-form gauge potential, one noteworthy feature of this solution is that the electromagnetic field strength carries both electric and magnetic components. Moreover, if we try to compute the electric and magnetic charges the results will depend on the radius of the $2$-sphere on which we integrate (see also \cite{Karlovini}). However, if we take the limit in which the $2$-sphere is pushed to infinity we find that the magnetic charge vanishes and the solution is purely electric with charge $q$.

We are now ready to present the general class of electrically-charged
Taub-NUT metrics in even dimensions $D=2d+2$. These spaces are constructed
as complex line bundles over an Einstein-K\"{a}hler space $M$, with
dimension $2d$ and metric $g_{M}$. The metric ansatz that we use is the
following: 
\begin{equation}
ds_{D}^{2}=-F(r)(dt-2N\mathcal{A})^{2}+F^{-1}(r)dr^{2}+(r^{2}+N^{2})g_{M}
\end{equation}
Here $\mathcal{J}=d\mathcal{A}$ is the K\"{a}hler form for the
Einstein-Kahler space $M$ and we use the normalisation such that the Ricci
tensor of the Einstein-K\"{a}hler space $M$ is $R_{ab}=\delta g_{ab}$.

Motivated by the known four-dimensional solution we shall make the following
ansatz for the electromagnetic gauge potential: 
\begin{eqnarray}
A&=&-\sqrt{\frac{D-2}{2}}\frac{qr}{(r^2+N^2)^{\frac{D-2}{2}}}(dt-2N\mathcal{A%
})
\end{eqnarray}

Then the general solution to Einstein's field equations with cosmological
constant $\lambda =\pm (D-1)/l^{2}$ is given by: 
\begin{eqnarray}
F(r)&=&\frac{r}{(r^{2}+N^{2})^{\frac{D-2}{2}}}\bigg[ \int\limits^{r}\left(
\delta\mp \frac{D-1}{l^{2}}(s^{2}+N^{2})\right) \frac{(s^{2}+N^{2})^{\frac{%
D-2}{2}}}{s^{2}}ds-2m\bigg]  \notag \\
&&+q^2\frac{(D-3)r^2+N^2}{(r^2+N^2)^{D-2}}  \label{GenF}
\end{eqnarray}
 
 As in the $4$-dimensional case the electromagnetic field strength has both electric and magnetic components. If we try to compute the electric and magnetic charges we obtain again results that depend on the radial coordinate $r$. However, if we push the integration surfaces to infinity the magnetic charge will vanish leaving us only with an effective electric charge.
 
As an example of this general solution, let us assume that the $(D-2)$-dimensional base space in our construction is a product of $d$ factors, $M=M_{1}\times \dots \times M_{d}$ where $M_{i}$ are two
dimensional Einstein-K\"{a}hler spaces or more generally $CP^n$ factors. In particular, we can use the sphere $S^{2}$, the torus $T^{2}$ or the hyperboloid $H^{2}$ as factor spaces. It is then easy to see that if $q=0$ we recover the previously known higher dimensional Taub-NUT solutions studied in refs. \cite{Page1,Bais,Robinson,Awad,Lorenzo}. On the other hand, if $N=0$ then we recover the topological Reissner-Nordstr\"{o}m-AdS solutions given in \cite{Chamblin,eugen}.

\section{Regularity conditions}

Scalar curvature singularities have the possibility of manifesting themselves only in the Euclidean sections. These are simply obtained by the analytic continuations $t\rightarrow i\tau$, $N\rightarrow in$ and $q\rightarrow iq$, and can be classified by the dimensionality of the fixed point sets of the Killing vector $\xi =\partial /\partial \tau$\ that generates a $U(1)$ isometry group. In four dimensions, the Killing vector that corresponds to the coordinate that parameterizes the fibre $S^{1}$ can have a zero-dimensional fixed point set (we speak about a `NUT' solution in this case) or a two-dimensional fixed point set (referred to as a `bolt' solution). The classification in higher dimensions can be done in a similar
manner. If this fixed point set dimension is $\left( D-2\right) $ the solution is called a Bolt solution; if the dimensionality is less than this then the solution is called a NUT solution. Notice that if $D>4$ then NUTs
with larger dimensionality can exist \cite{csrm,Page1}. In general, fixed point sets need not exist; indeed there are parameter ranges of NUT-charged asymptotically $dS$ spacetimes that have no Bolts \cite{Anderson}.

The singularity analysis performed here is a direct application of the one given in \cite{Page}. In order to extend the local metrics presented above to global metrics on non-singular manifolds the idea is to turn all the singularities appearing in the metric into removable coordinate singularities. For generic values of the parameters the singularities are not removable, corresponding to conical singularities in the manifold. We are mainly interested in the case of a compact Einstein-K$\ddot{a}$hler manifold $M$. Generically the K\"{a}hler form $\mathcal{J}$ on $M$ can be equal to $d\mathcal{A}$ only locally and we need to use a number of overlapping coordinate patches to cover the whole manifold. Now, in order to render the $1$-form $d\tau -2n\mathcal{A}$ well-defined we need
to identify $\tau $ periodically. We will require the period of $\tau $ to be given by: 
\begin{equation}
\beta =\frac{4\pi np}{k\delta }\label{beta}
\end{equation}
where $k$ is a positive integer, while $p$ is a non-negative integer, defined as the integer such that the first Chern class, $c_{1}$, evaluated on $H_{2}(M)$ is $\mathbf{Z}\cdot p$, \textit{i.e.} the integers divisible by $p$. Among all the Einstein-K\"{a}hler manifolds the integer $p$ is maximised in $CP^{q}$, for which $p=q+1$ \cite{Page}. It is also necessary to eliminate the singularities in the metric that appear as $r$ is varied over $M$. The critical points are to the so-called endpoint values of $r$:
these are the values for which the metric components become zero or infinite. For a complete manifold $r$ must range between two adjacent endpoints and we must eliminate the conical singularities (if any) which occur at these points. The finite endpoins occur at $r=\pm n$ or at the simple zeros of $F_{E}(r)$. Quite generally, when the electrical charge $q$ is zero, $r=\pm n$ give the location of curvature singularities unless $F_{E}=0$ there as well. By contrast with the uncharged case, it turns out
that if $q\neq 0$ then the curvature singularities at $r=\pm n$ cannot be eliminated for any choices of the parameters. This can be seen from the fact that $F_E(r)$ diverges badly when $r\ra \pm n$ for any values of $m$ and the components of the curvature tensor will diverge there as well. Therefore, in order to obtain non-singular Euclidian sections we have to restrict the range of the radial coordinate such that the values $r=\pm n$ are avoided. We should then restrict our attention to simple roots $r_{0}$ of $F_{E}(r)$ different from $\pm n$. In general, to eliminate a conical singularity at a root $r_{0}$ of $F_{E}(r)$ we must restrict the periodicity of $\tau $ to be given by: 
\begin{equation*}
\beta =\frac{4\pi }{|F_{E}^{\prime }(r_{0})|}
\end{equation*}
and this will generally impose a restriction on the values of the parameters once we match it with ($\ref{beta}$). This condition will also fix the location of the bolt, which will be given by a solution of the equation: 
\begin{equation}
npq^{2}\big[(D-3)^{2}r_{0}^{4}-2n^{2}r_{0}^{2}+n^{4}\big]+\bigg[np\left(
\delta -\lambda (r_{0}^{2}-n^{2})\right) -k\delta r_{0}\bigg](r_{0}^{2}-n^{2})^{D-1}=0\label{Genrb}
\end{equation}

For compact manifolds the radial coordinate takes values between two finite endpoints and we have to impose this constraint at both endpoints. If the manifold is noncompact then the cosmological constant must be non-positive and the radial coordinate takes values between one finite endpoint $r_{0}$
and one infinite endpoint $r_{1}=\infty $. Since for our asymptotically locally $(A)dS$ or flat solutions the infinite endpoints are not within a finite distance from any points $r\neq r_{1}$ there is no regularity
condition to be imposed at $r_{1}$. In this case the only regularity conditions are that $F_{E}(r)>0$ for $r\geq r_{0}$ and $\beta =\frac{4\pi }{|F_{E}^{\prime }(r_{0})|}$ to be satisfied. The only way to avoid a
curvature singularity at $r=n$ is to restrict the values of the radial coordinate such that $r\geq r_{0}>n$, \textit{i.e.} the only non-singular Taub-NUT-RN spaces are the TNRN-bolt solutions\footnote{We are focussing here on non-compact manifolds. For compact manifolds we could restrict the range of the radial coordinate $r$ to the interval between two adjacent roots $r_{1}\leq r\leq r_{2}$ of $F_{E}(r)$ such that the values $\pm n$ are avoided.}.

\section{The Taub-NUT-RN solution in six dimensions}

As an illustration of the general analysis performed in the previous
section, in this section we shall look more closely at a six-dimensional
Taub-NUT-RN solution constructed over the four-dimensional base $S^{2}\times
S^{2}$. Performing the analytical continuations $t\rightarrow i\tau $, $%
N\rightarrow in$ and $q\rightarrow iq$, the metric in the Euclidian sector
can be written in the form: 
\begin{eqnarray}
ds^{2} &=&F_{E}(r)\left( d\tau -2n\cos \theta _{1}d\phi _{1}-2n\cos \theta
_{2}d\phi _{2}\right) ^{2}+\frac{dr^{2}}{F_{E}(r)}  \notag \\
&&+(r^{2}-n^{2})(d\theta _{1}^{2}+\sin ^{2}\theta _{1}d\phi
_{1}^{2})+(r^{2}-n^{2})(d\theta _{2}^{2}+\sin ^{2}\theta _{2}d\phi _{2}^{2})
\end{eqnarray}%
where the function $F_{E}(r)$, respectively the $1$-form potential $A$ are
given by: 
\begin{eqnarray}
F_{E}(r) &=&\frac{%
3r^{6}+(l^{2}-15n^{2})r^{4}+3n^{2}(15n^{2}-2l^{2})r^{2}-6ml^{2}r+3n^{4}(5n^{2}-l^{2})%
}{3(r^{2}-n^{2})^{2}l^{2}}  \notag \\
&&-\frac{q^{2}\left( 3r^{2}-n^{2}\right) }{(r^{2}-n^{2})^{4}}  \notag \\
A &=&-\frac{\sqrt{2}qr}{(r^{2}-n^{2})^{2}}\left( d\tau -2n\cos \theta
_{1}d\phi _{1}-2n\cos \theta _{2}d\phi _{2}\right)
\end{eqnarray}%
Regularity of the $1$-form $d\tau -2n\mathcal{A}$ forces the periodicity of
the Euclidian time to be $\frac{8\pi n}{k}$, for some integer $k$. As
mentioned in the previous section, the NUT solution is singular thence we
restrict our attention directly to the Bolt solutions. These solutions correspond to a
four-dimensional fixed-point set located at a simple root $r_{b}$ of $%
F_{E}(r)$ and we restrict the values of the radial coordinate such that $%
r\geq r_{b}>n$. The periodicity of $\tau $ is given by $\frac{8\pi n}{k}$
and we have to match it with the periodicity obtained by eliminating the
conical singularities at $r_{b}$. This will fix the location of the bolt as
given by a root of 
\begin{equation*}
2nq^{2}(9r_{b}^{4}-2n^{2}r_{b}^{2}+n^{4})+\bigg[2n\left( 1+\frac{5}{l^{2}}%
(r_{b}^{2}-n^{2})\right) -kr_{b}\bigg](r_{b}^{2}-n^{2})^{5}=0
\end{equation*}%
As this is a polynomial equation of rank $12$, an analytical solution for $%
r_{b}$ seems out of the question.

Finally, the value of the mass parameter is given by: 
\begin{eqnarray}
m_{b} &=&\frac{%
3r_{b}^{10}+(l^{2}-21)r_{b}^{8}+n^{2}(78n^{2}-8l^{2})r_{b}^{6}+10n^{4}(l^{2}-9n^{2})r_{b}^{4}+(15n^{8}-9q^{2}l^{2})r_{b}^{2}%
}{6(r_{b}^{2}-n^{2})^{2}l^{2}r_{b}}  \notag \\
&&-\frac{3n^{2}(n^{6}l^{2}-q^{2}l^{2}-5n^{8})}{%
6(r_{b}^{2}-n^{2})^{2}l^{2}r_{b}}.  \notag
\end{eqnarray}%
Generically there is a curvature singularity at $r=n$, which is simply
avoided if we restrict the range of the radial coordinate such that $r\geq
r_{b}>n$. If $q=0$ we recover the six-dimensional cosmological Taub-NUT
solution over the base space $S^{2}\times S^{2}$, which was discussed in %
\cite{Awad,csrm}. If $n=0$ the solution reduces to the topological
Reissner-Nordstr\"{o}m solution whose horizon topology is $S^{2}\times
S^{2}$.

Similar results are obtained for a fibration over $CP^2$. The only difference appears in the periodicity of $\tau$, which has to be now $12\pi n/k$ and this will also modify the equation for $r_b$ (as it can be read from the general expression (\ref{Genrb}) with $p=3$ and $\delta=1$). Unlike the uncharged case, the NUT solution is singular as there will be a curvature singularity at $r=n$.

\section{Conclusions}

In this paper we presented new families of higher dimensional solutions of
the sourceless Einstein-Maxwell field equations with or without cosmological
constant. These solutions are constructed as radial extensions of circle
fibrations over even dimensional spaces that can be in general products of
Einstein-K\"{a}hler spaces.

We have given the Lorentzian form of the solutions. However in order to
understand the singularity structure of these spaces, we have concentrated
mainly on their Euclidian sections -- recognising that the Lorentzian
versions are singularity-free -- apart from quasi-regular singularities \cite%
{Konk}. These correspond to the end-points of incomplete and inextensible
geodesics that spiral infinitely around a topologically closed spatial
dimension. However, since the Riemann tensor and all its derivatives remain
finite in all parallelly propagated orthonormal frames, we take the point of
view that these represent some of mildest of types of singularities and we
ignore them when discussing the singularity structure of the Taub-NUT
solutions.

In general the Euclidean section is simply obtained using the analytic
continuations $t\rightarrow i\tau$, $N\rightarrow in$ and $q\rightarrow iq$.
Generically the Taub-NUT solutions present themselves in two classes: `nuts'
and 'bolts'. While in the uncharged case there can exist NUTs with
intermediate dimensionality \cite{csrm,Page1,TNletter}, for our NUT
solutions the fix-point set has always dimension $0$ and it is singular.
Indeed, we found that at the NUT location there always exists a curvature
singularity that cannot be eliminated for any choice of the parameters. This
is clearly in contrast with the uncharged case: in absence of the electrical
charge, there are NUT solutions that are non-singular -- for example, if we
use $M=CP^q$ then for an appropriate choice of the mass parameter we find
that there is no curvature singularity at $r=n$. The only regular charged
solutions are then the Bolt metrics. The regularity conditions require us to
fix the periodicity of the Euclidian time $\tau$. This periodicity is
determined in two ways and by matching the two obtained values we get
restrictions on the values of the parameters in our solutions. These
restrictions will also fix the location of the bolt. However, as in $D$%
-dimensions the equation that we have to solve is a polynomial equation of
rank $2D$, there is no chance to obtain the bolt location in a closed
analytical form.

Finally, we note that for the present solutions the boundary is generically
a circle-fibration over base spaces that, being obtained from products of
general Einstein-K\"{a}hler manifolds, can have exotic topologies. It
would be interesting to see if they could be used as test-grounds for the $
AdS/CFT$ correspondence and more generally in context of gauge/gravity
dualities. We leave a more detailed study of this subject for future work.

\medskip
{\it Note added} After the completion of this manuscript we learned about the paper \cite{Awad} that presents similar higher-dimensional solutions.

\medskip

{\Large Acknowledgements}

This work was supported by the Natural Sciences and Engineering Council of
Canada.

\end{document}